\documentclass[12pt]{article}
\usepackage{amsmath}
\usepackage{amsfonts}
\usepackage{amssymb}
\usepackage{graphicx}
\usepackage{breqn}
\usepackage{array}
\usepackage{tabu}
\usepackage{tabularx,ragged2e,booktabs,caption}
\usepackage{float}
\usepackage{verbatim}
\usepackage{hyperref}
\usepackage{cite}
\usepackage[numbers,square,compress]{natbib}
\newcommand{\bea}{\begin{eqnarray}}
\newcommand{\eea}{\end{eqnarray}}
\newcommand{\be}{\begin{equation}}
\newcommand{\ee}{\end{equation}}
\begin{document}
\thispagestyle{empty}
\begin{flushright}
September 2016
\end{flushright}
\vspace*{2cm}
\begin{center}
{\Large \bf Stationary Charged Scalar Clouds around }\\	\vskip.3cm
{\Large \bf Black Holes in String Theory}\\
\vspace*{1.8cm} {\large Canisius Bernard$^{\dagger ,}$\footnote{\texttt{canisius.bernard@gmail.com}}\\
\vspace{.5cm} {\em $^{\dagger}$Center for Theoretical Physics,\\ Department of Physics, Parahyangan Catholic University,\\ Bandung 40141, Indonesia}\\
\vspace{.15cm} \vspace{1cm} ABSTRACT}
\end{center}

It was reported that Kerr-Newman black holes can support linear charged scalar fields in their exterior regions. These stationary massive charged scalar fields can form bound states, which are called stationary scalar clouds. In this paper, we show that Kerr-Sen black holes can also support stationary massive charged scalar clouds by matching the near- and far-region solutions of the radial part of the Klein-Gordon wave equation. We also review stationary scalar clouds within the background of static electrically charged black hole solutions in the low-energy limit of heterotic string field theory, namely, the Gibbons-Maeda-Garfinkle-Horowitz-Strominger black holes.
\begin{description}
\item[PACS numbers:] 04.70.Bw, 04.60.Cf, 04.50.Kd, 04.20.Jb
\item[Keywords:] scalar cloud, black hole, string theory
\end{description}
%%%%%%%%%%%%%%%%%%%%%%%%%%%%%%%%%%%%%%%%%%%%%%%%%%%%%%%%%%%%%%%
\vfill \setcounter{page}{0} \setcounter{footnote}{1}
\newpage
\section{Introduction}
\label{sec:intro} 
The uniqueness theorem states that all black hole solutions of the Einstein-Maxwell equation of gravitation and electromagnetism can be completely characterized by only three parameters: mass, electric charge, and angular momentum \cite{Chrusciel:2012jk,Herdeiro:2015waa}. As we know, the Kerr-Newman black hole is a solution for a rotating and charged black hole in the Einstein-Maxwell field equation. However, rotating and charged black hole solutions can also be found in other theories. For example, there is also a rotating and charged black hole solution known as a Kerr-Sen black hole, which is a solution to a set of classical equations of motion arising in the low energy limit of heterotic string field theory. In 1992, Sen \cite{Sen:1992ua} was the first to derive this solution by transforming the Kerr metric into a Kerr-Sen one. As we know, both Kerr-Newman and Kerr-Sen black holes have quite similar physical properties because both are rotating and charged black hole solutions. The motivation for studying Kerr-Sen black holes is perhaps that the Universe does not precisely follow Einstein-Maxwell theory, but rather a more complex one, namely, string theory. If this is the case, the expected rotating and charged black hole solution would be the Kerr-Sen solution instead of the Kerr-Newman one.\\
\indent Hod proposed that Kerr-Newman black holes can support linear charged scalar fields in the near-extremal regions \cite{Hod:2014baa}. These bound states are the stationary scalar configurations in the black hole backgrounds, which are regular at the horizon and outside it. They are called stationary scalar clouds \cite{Hod:2014baa,Herdeiro:2014goa}. The importance of these clouds is that they suggest the existence of hairy black holes in the fully nonlinear regime, i.e., considering the full Einstein-scalar theory. This was first shown in \cite{Herdeiro:2014goa}, complemented by \cite{Herdeiro:2015gia}, and conjectured as a general mechanism in \cite{Herdeiro:2014ima}. The existence of a stationary bound state of a charged massive scalar field in black hole backgrounds should have an asymptotically exponentially decaying radial solution behavior \cite{Benone:2014ssa}. By analyzing the radial equation (\ref{SV04}), we found that this asymptotic radial solution matched this requirement well \cite{Siahaan:2015xna},
\be \label{IN01}
R_{lm}\left( r \right) \approx \left\{ \begin{array}{l}
{e^{ - i\left( {\omega  - {\omega _c}} \right){r_*}}} ~~~~~{\rm{ for }}~r \to {r_ + },\\
{e^{ - i{r_*}\sqrt {{\omega ^2} - {\mu ^2}} }} ~~~{\rm{ for }}~r \to \infty ,{\rm{ }}
\end{array} \right.
\ee 
where $\omega_c$ is the critical frequency
\be \label{IN02}
{\omega _c} \equiv m{\Omega _H} + q{\Phi _H}.
\ee
Here, $r_+$ is the radial coordinate of the outer horizon defined in Eq. (\ref{KSG07}), ${\Omega _H}$ is the angular velocity, and ${\Phi _H}$ is the electrostatic potential defined in Eqs. (\ref{KSG15}) and (\ref{KSG16}). The ``tortoise'' radial coordinate $r_*$ is defined by
\be \label{IN03}
\frac{{d{r_*}}}{{dr}} = \frac{{{\Delta _{KS}} + 2Mr}}{{{\Delta _{KS}}}}.
\ee
The critical frequency $\omega_c$ in Kerr-Sen black hole backgrounds is the same as the one found in the Kerr-Newman case. It is not surprising because they both describe the spacetime outside of charged and rotating black holes \cite{Siahaan:2015xna}. In this paper, we study whether there is a scalar cloud surrounding a Kerr-Sen black hole.\\
\indent The paper is organized as follows. In Sec. \ref{sec:kerr-sen} we review the Kerr-Sen geometry and the corresponding Hawking temperature, angular velocity, and electrostatic potential. In Sec. \ref{sec:angular-radial} we separate the scalar field equation on Kerr-Sen geometry into the angular and radial parts. In Sec. \ref{sec:far} we study and derive the radial solution in the far region, defined by $x\gg\tau$. Then in Sec. \ref{sec:near} we derive the radial solution in the near region defined by $x\ll 1$, and the solution in the matching region defined by $\tau\ll x\ll1$ is given in Sec. \ref{sec:matching}. Finally, the stationary bound state of charged massive scalar fields in the Kerr-Sen black hole is given in Sec. \ref{sec:bound-state}. In Sec. \ref{sec:GMGHS} we review the Gibbons-Maeda-Garfinkle-Horowitz-Strominger (GMGHS) spacetime and stationary charged scalar clouds in GMGHS spacetime. In the last section we give our conclusions. In this paper we use units in which $G = c = \hbar = 1$.
%%%%%%%%%%%%%%%%%%%%%%%%%%%%%%%%%%%%%%%%%%%%%%%%%%%%%%%%%%%%%%%
\section{Kerr-Sen Geometry}
\label{sec:kerr-sen}
In 1992, Sen derived a four-dimensional charged and rotating black hole solution in the low energy limit of heterotic string field theory. The string theory effective action in four dimensions is given by \cite{Siahaan:2015xna,Ghezelbash:2012qn,Siahaan:2015ljs}
\be \label{KSG01}
S = \int {{d^4}x\sqrt { - g} } {e^{ - \tilde \Phi }}\left( {R - \frac{1}{8}{F_{\mu \nu }}{F^{\mu \nu }} + {g^{\mu \nu }}{\partial _\mu }\tilde \Phi {\partial _\nu }\tilde \Phi  - \frac{1}{{12}}{H_{\kappa \lambda \mu }}{H^{\kappa \lambda \mu }}} \right),
\ee
where $g$ is the determinant of the tensor metric $g_{\mu\nu}$, $R$ is the scalar curvature, $\tilde \Phi$ is the dilaton field, $F_{\mu\nu}$ is the field-strength tensor
\be \label{KSG02}
{F_{\mu \nu }} = {\partial _\mu }{A_\nu } - {\partial _\nu }{A_\mu },
\ee
and ${H_{\kappa \lambda \mu }}$ is the third-rank tensor field
\be \label{KSG03}
{H_{\kappa \mu \nu }} = {\partial _\kappa }{B_{\mu \nu }} + {\partial _\nu }{B_{\kappa \mu }} + {\partial _\mu }{B_{\nu \kappa }} - \frac{1}{4}\left( {{A_\kappa }{F_{\mu \nu }} + {A_\nu }{F_{\kappa \mu }} + {A_\mu }{F_{\nu \kappa }}} \right).
\ee
Note that $B_{\nu\sigma}$ is a second-rank antisymmetric tensor gauge field. Sen applied a transformation to the Kerr solution, known as a solution to the vacuum Einstein equation, to obtain the charged rotating black hole solution in the theory (\ref{KSG01}), known as the Kerr-Sen solution. In Boyer-Lindquist coordinates ($t,r,\theta,\phi$), the Kerr-Sen metric in the Einstein frame can be read as \cite{Siahaan:2015xna,Wu:2003qb}
\bea \nonumber
d{s^2} &=&  - \left( {1 - \frac{{2Mr}}{{{\rho ^2}}}} \right)d{t^2} + {\rho ^2}\left( {\frac{{d{r^2}}}{{{\Delta _{KS}}}} + d{\theta ^2}} \right) - \frac{{4Mra}}{{{\rho ^2}}}{\sin ^2}\theta dtd\phi  \\\label{KSG04}&\quad&+ \left( {{\rho ^2} + {a^2}{{\sin }^2}\theta  + \frac{{2Mr{a^2}{{\sin }^2}\theta }}{{{\rho ^2}}}} \right){\sin ^2}\theta d{\phi ^2},
\eea
where,
\bea \label{KSG05}
\Delta _{KS}  &=& {r^2} + 2(b - M)r + {a^2},\\ \label{KSG06}
\rho^2  &=& {r^2} + 2br + {a^2}{\cos ^2}\theta ,\\ \label{KSG07}
{r_ \pm } &=& M - b \pm \sqrt {{{\left( {M - b} \right)}^2} - {a^2}} ,\\ \label{KSG08}
b &=& \frac{{{Q^2}}}{{2M}},
\eea
where $r_+$ is the outer horizon and $r_-$ is the inner horizon. The nonvanishing components of the Kerr-Sen contravariant tensor metric in the Einstein frame are
\bea \nonumber
{g^{tt}} &=& \frac{{{\Delta _{KS}}{a^2}{{\sin }^2}\theta  - {{\left( {{r^2}+2br + {a^2}} \right)}^2}}}{{{\Delta _{KS}}\rho^2 }},~{g^{rr}} = \frac{{{\Delta _{KS}}}}{\rho^2 },~\\\label{KSG09}
{g^{\theta \theta }} &=& \frac{1}{\rho^2 },~{g^{\phi \phi }} = \frac{{{\Delta _{KS}} - {a^2}{{\sin }^2}\theta }}{{{\Delta _{KS}}\rho^2 {{\sin }^2}\theta }},~\\\nonumber
{g^{t\phi }} &=& {g^{\phi t}} =  - \frac{2Mar}{{{\Delta _{KS}}\rho^2 }},
\eea
with $\sqrt{-g}=\rho^2 \sin(\theta)$. The metric (\ref{KSG04}) describes a black hole with mass $M$, charge $Q$, and angular momentum $J=Ma$. The solutions for nongravitational fields are 
\bea \label{KSG10}
\tilde \Phi  &=&  - \frac{1}{2}\ln \frac{{{\rho ^2}}}{{{r^2} + {a^2}{{\cos }^2}\theta }},\\ \label{KSG11}
{A_t} &=&  - \frac{{Qr}}{{{\rho ^2}}},\\ \label{KSG12}
{A_\phi } &=& \frac{{Qra{{\sin }^2}\theta }}{{{\rho ^2}}},\\ \label{KSG13}
{B_{t\phi }} &=& \frac{{bra{{\sin }^2}\theta }}{{{\rho ^2}}},
\eea 
and the related Hawking temperature, angular velocity, and electrostatic potential at the horizon are given by
\bea \label{KSG14}
{T_H} &=& \frac{{{r_ + } - {r_ - }}}{{8\pi M{r_ + }}},\\ \label{KSG15}
{\Omega _H} &=& \frac{a}{{2M{r_ + }}},\\ \label{KSG16}
{\Phi _H} &=& \frac{Q}{{2M}}.
\eea
\indent Setting $b=0$, solutions for nongravitational fields (\ref{KSG10})--(\ref{KSG13}) vanish, and therefore the Kerr-Sen metric (\ref{KSG04}) reduces to the Kerr metric. Instead, turning off the rotational parameter $a$ followed by a coordinate transformation $r\rightarrow r-Q^2/M$ transforms the Kerr-Sen solution into the GMGHS solution which describes a static electrically charged black hole in string theory \cite{Gibbons:1987ps,Garfinkle:1990qj}.
%%%%%%%%%%%%%%%%%%%%%%%%%%%%%%%%%%%%%%%%%%%%%%%%%%%%%%%%%%%%%%%
\section{Separation of Variables}
\label{sec:angular-radial}
Let us conceive a massive charged scalar particle field $\Phi$ outside of a Kerr-Sen black hole with mass $\mu$ and charge $q$ that obey the subsequent Klein-Gordon wave equation \cite{Konoplya:2013rxa}
\be \label{SV01}
\frac{1}{{\sqrt { - g} }}{\partial _\alpha }\left( {{g^{\alpha \beta }}\sqrt { - g} {\partial _\beta }\Phi } \right) - 2iq{A_\alpha }{g^{\alpha \beta }}{\partial _\beta }\Phi  - {q^2}{g^{\alpha \beta }}{A_\alpha }{A_\beta }\Phi  - {\mu ^2}\Phi  = 0.
\ee
To solve the equation above, as usual we can use the ansatz \cite{Brill:1972xj}
\be \label{SV02}
\Phi  = \sum\limits_{l,m} {{\Phi _{lm}}} =\sum\limits_{l,m} {e^{i\left( {m\phi  - \omega t} \right)}}R_{lm}\left( r \right)S_{lm}\left( \theta  \right),
\ee
where $\omega$ is the frequency of the wave field, $l$ is the spheroidal harmonic index, and $m$ is the azimuthal harmonic index. Substituting Eq. (\ref{SV02}) into (\ref{SV01}) gives us two separated equations, specifically the angular part
\be \label{SV03}
\frac{1}{{\sin \theta }}\frac{d}{{d\theta }}\left( {\sin \theta \frac{dS_{lm}\left( \theta  \right)}{{d\theta }}} \right) + \left[ {\lambda_{lm} + {a^2}\left( {{\mu ^2} - {\omega ^2}} \right) - {a^2}{{\cos }^2}\theta \left( {{\mu ^2} - {\omega ^2}} \right) - \frac{{{m^2}}}{{{{\sin }^2}\theta }}} \right]S_{lm}\left( \theta  \right) = 0,
\ee
and the radial part
\be \label{SV04}
\frac{d}{{dr}}\left( {\Delta _{KS}\frac{dR_{lm}\left( r \right)}{{dr}}} \right) + \left[ {\frac{{{G^2}}}{\Delta _{KS}} - {\mu ^2}\left( {{r^2}+ 2br + {a^2}} \right) + 2am\omega  - \lambda_{lm}} \right]R_{lm}\left( r \right) = 0,
\ee
with
\be \label{SV05}
G = \omega \left( {{r^2} + 2br + {a^2}} \right) - qQr - am.
\ee
The coupling constant $\lambda_{lm}$ may be expanded as a
power series \cite{Dolan:2007mj}
\be \label{SV06}
{\lambda _{lm}} + {a^2}\left( {{\mu ^2} - {\omega ^2}} \right) = l\left( {l + 1} \right) + \sum\limits_{k = 1}^\infty  {{c_k}{a^{2k}}{{\left( {{\mu ^2} - {\omega ^2}} \right)}^k},}
\ee
where coefficients $c_k$ are given in \cite{Abra}.\\
\indent The charged massive scalar field with the frequency $\omega=\omega_c$ represents the stationary bound state of the charged massive scalar field in Kerr-Sen spacetime. Following \cite{Hod:2014baa}, we outline the new dimensionless variables
\bea \label{SV07}
x &\equiv& \frac{{r - {r_ + }}}{{{r_ + }}},\\\label{SV08}
\tau  &\equiv& \frac{{{r_ + } - {r_ - }}}{{{r_ + }}},\\\label{SV09}
k &\equiv& 2{\omega _c}{r_ + } - qQ.
\eea
Therefore, the radial Teukolsky equation (\ref{SV04}) can be rewritten in the form
\be \label{SV10}
x\left( {x + \tau } \right)\frac{{{d^{2}R(x)}}}{{d{x^2}}} + \left( {2x + \tau } \right)\frac{dR(x)}{{dx}} + \mathcal{V}R\left( x \right) = 0,
\ee
where
\bea \label{SV11}
\mathcal{V} &=& \frac{{{G^2}}}{{{r_ + }^2x\left( {x + \tau } \right)}} - \lambda  + 2am{\omega _c} - {\mu ^2}\left[ {{r_ + }^2{{\left( {x + 1} \right)}^2} + 2b{r_ + }\left( {x + 1} \right) + {a^2}} \right],\\ \label{SV12}
G &=& r_ + ^2{\omega _c}{x^2} + (2b{\omega _c}+k){r_ + } x.
\eea
%%%%%%%%%%%%%%%%%%%%%%%%%%%%%%%%%%%%%%%%%%%%%%%%%%%%%%%%%%%%%%%
\section{Far-Region Analysis}
\label{sec:far}
Now we analyze the radial Teukolsky equation (\ref{SV10}) within the region $x \gg \tau $. In this far region, Eq. (\ref{SV10}) can approximately be expressed as
\be \label{FR01}
{x^2}\frac{{{d^{2}R(x)}}}{{d{x^2}}} + 2x\frac{dR(x)}{{dx}} + {\mathcal{V}_{far}}R\left( x \right) = 0,
\ee
with $\mathcal{V}\equiv{\mathcal{V}_{far}}$,
\be \label{FR02}
{\mathcal{V}_{far}} \equiv {\left( {\left( {2b + r_ + ^{}x} \right){\omega _c} + k} \right)^2} - \lambda  + 2am{\omega _c} - {\mu ^2}\left[ {{r_ + }^2{{\left( {x + 1} \right)}^2} + 2b{r_ + }\left( {x + 1} \right) + {a^2}} \right].
\ee
The solution of Eq. (\ref{FR01}) is \cite{Abra}
\bea \nonumber
R\left( x \right) &=& {\mathcal{C}_1} \times {\left( {2\tilde\varepsilon } \right)^{\frac{1}{2} + \tilde\beta }}{x^{ - \frac{1}{2} + \tilde\beta }}{e^{ - \tilde\varepsilon x}}M\left( {\frac{1}{2} + \tilde\beta  - \tilde\kappa ,1 + 2\tilde\beta ;2\tilde\varepsilon x} \right)\\\label{FR03}
&\quad& +~{\mathcal{C}_2} \times {\left( {2\tilde\varepsilon } \right)^{\frac{1}{2} - \tilde\beta }}{x^{ - \frac{1}{2} - \tilde\beta }}{e^{ - \tilde\varepsilon x}}M\left( {\frac{1}{2} - \tilde\beta  - \tilde\kappa ,1 - 2\tilde\beta ;2\tilde\varepsilon x} \right),
\eea
with
\be \label{FR04}
\tilde\varepsilon  \equiv {r_ + }\sqrt {{\mu ^2} - {\omega _c}^2}.
\ee
In Eq. (\ref{FR03}) above, $M(a,b;z)$ is the Whittaker function, and ${\mathcal{C}_1},~{\mathcal{C}_2}$ are some normalization constants. In addition,
\bea \label{FR05}
{\tilde\beta ^2} &\equiv& \frac{1}{4} + \lambda  - 2am{\omega _c} - {\left( {2b{\omega _c} + k} \right)^2} + {\mu ^2}\left( {{r_ + }^2 + 2b{r_ + } + {a^2}} \right),\\\label{FR06}
\tilde\kappa  &\equiv& \frac{{{\omega _c}\left( {2b{\omega _c} + k} \right) - {\mu ^2}\left( {b + {r_ + }} \right)}}{{\sqrt {{\mu ^2} - {\omega _c}^2} }}.
\eea 
%%%%%%%%%%%%%%%%%%%%%%%%%%%%%%%%%%%%%%%%%%%%%%%%%%%%%%%%%%%%%%%
\section{Near-Region Analysis}
\label{sec:near}
Next we analyze the radial Teukolsky equation (\ref{SV10}) within the region $x\ll1$. In this near region, the effective potential (\ref{SV11}) approaches
\bea \label{NR01}
{\mathcal{V}_{near}} &\equiv& \frac{{{{\left( {2b{\omega _c} + k} \right)}^2}x}}{{\left( {x + \tau } \right)}} - \lambda  + 2am{\omega _c} - {\mu ^2}\left[ {{r_ + }^2 + 2b{r_ + } + {a^2}} \right].
\eea
The solution of Eq. (\ref{SV10}) with $\mathcal{V}\equiv\mathcal{V}_{near}$ is \cite{Abra}
\be \label{NR02}
{R}\left( x \right) = {\left( {\frac{x}{\tau } + 1} \right)^{{-i (2b{\omega _c}+k)}}}~_2{F_1}\left( {\frac{1}{2} + \tilde\beta  - i {(2b{\omega _c}+k)} ,\frac{1}{2} - \tilde\beta  - i {(2b{\omega _c}+k)} ;1; - \frac{x}{\tau }} \right),
\ee 
where $_2{F_1}\left( {a,b;c;z} \right)$ is the hypergeometric function and $\tilde\beta$ is the same as Eq. (\ref{FR05}).
%%%%%%%%%%%%%%%%%%%%%%%%%%%%%%%%%%%%%%%%%%%%%%%%%%%%%%%%%%%%%%%
\section{Matching-Region Analysis}
\label{sec:matching}
Consider the following condition for near-extremal black holes:
\be \label{MR00}
\tau \ll 1.
\ee
For near-extremal Kerr-Sen black holes with $\tau \ll 1$, there is a matching region $\tau\ll x\ll1$; therefore, the Eqs. (\ref{FR03}) and (\ref{NR02}) can be matched. For $x\ll1$, the limit of Eq. (\ref{FR03}) is given by \cite{Hod:2014baa,Hartman:2009nz}
\be \label{MR01}
R \to {\mathcal{C}_1} \times {\left( {2\tilde\varepsilon } \right)^{\frac{1}{2} + \tilde\beta }}{x^{ - \frac{1}{2} + \tilde\beta }} + {\mathcal{C}_2} \times {\left( {2\tilde\varepsilon } \right)^{\frac{1}{2} - \tilde\beta }}{x^{ - \frac{1}{2} - \tilde\beta }}.
\ee
For $x\gg\tau$, the limit of Eq. (\ref{NR02}) is given by \cite{Hod:2014baa,Hartman:2009nz}
\be \label{MR02}
R \to {\mathcal{C}_3}\times{x^{ - \frac{1}{2} + \tilde\beta }} + {\mathcal{C}_4}\times{x^{ - \frac{1}{2} - \tilde\beta }},
\ee
where
\bea \label{MR03}
{\mathcal{C}_3} &=& {\tau ^{\frac{1}{2} - \tilde\beta  }}\frac{{\Gamma \left( {2\tilde\beta } \right)}}{{\Gamma \left( {\frac{1}{2} + \tilde\beta  - i(2b{\omega _c}+k)} \right)\Gamma \left( {\frac{1}{2} + \tilde\beta  + i {(2b{\omega _c}+k)} } \right)}},\\\label{MR04}
{\mathcal{C}_4} &=& {\tau ^{\frac{1}{2} + \tilde\beta }}\frac{{\Gamma \left( {-2\tilde\beta } \right)}}{{\Gamma \left( {\frac{1}{2} - \tilde\beta  - i(2b{\omega _c}+k)} \right)\Gamma \left( {\frac{1}{2} - \tilde\beta  + i {(2b{\omega _c}+k)} } \right)}}.
\eea
Matching (\ref{MR01}) and (\ref{MR02}) in this region, we find the two normalization constants $\mathcal{C}_{1},~\mathcal{C}_{2}$ of Eq. (\ref{MR01}),
\bea \label{MR05}
{\mathcal{C}_1} &=& {\tau ^{\frac{1}{2} - \tilde\beta  }}{\left( {2\tilde\varepsilon } \right)^{ - \left( {\frac{1}{2} + \tilde\beta } \right)}}\frac{{\Gamma \left( {2\tilde\beta } \right)}}{{\Gamma \left( {\frac{1}{2} + \tilde\beta  - i(2b{\omega _c}+k)} \right)\Gamma \left( {\frac{1}{2} + \tilde\beta  + i {(2b{\omega _c}+k)} } \right)}},\\\label{MR06}
{\mathcal{C}_2} &=& {\tau ^{\frac{1}{2} + \tilde\beta  }}{\left( {2\tilde\varepsilon } \right)^{ - \left( {\frac{1}{2} - \tilde\beta } \right)}}\frac{{\Gamma \left( {-2\tilde\beta } \right)}}{{\Gamma \left( {\frac{1}{2} - \tilde\beta  - i(2b{\omega _c}+k)} \right)\Gamma \left( {\frac{1}{2} - \tilde\beta  + i {(2b{\omega _c}+k)} } \right)}}.
\eea
The expansion for $x\to\infty$ of Eq. (\ref{FR03}) is given by \cite{Hod:2014baa,Hod:2012px}
\[R \to \left[{\mathcal{C}_1} \times {\left( {2{\tilde\varepsilon} } \right)^{\tilde\kappa} }{x^{ - 1 + {\tilde\kappa} }}{\left( { - 1} \right)^{ - \frac{1}{2} - {\tilde\beta}  + {\tilde\kappa} }}\frac{{\Gamma \left( {1 + 2{\tilde\beta}  } \right)}}{{\Gamma \left( {\frac{1}{2} + {\tilde\beta}   + {\tilde\kappa} } \right)}}\right.\]
\[\left.~+ {\mathcal{C}_2} \times {\left( {2{\tilde\varepsilon} } \right)^{\tilde\kappa} }{x^{ - 1 + {\tilde\kappa} }}{\left( { - 1} \right)^{ - \frac{1}{2} + {\tilde\beta}   + {\tilde\kappa} }}\frac{{\Gamma \left( {1 - 2{\tilde\beta}  } \right)}}{{\Gamma \left( {\frac{1}{2} - {\tilde\beta}   + {\tilde\kappa} } \right)}}\right]{e^{ - {\tilde\varepsilon} x}}\]
\be \label{MR07}
+ \left[ {{\mathcal{C}_1} \times {{\left( {2{\tilde\varepsilon} } \right)}^{ - {\tilde\kappa} }}{x^{ - 1 - {\tilde\kappa} }}\frac{{\Gamma \left( {1 + 2{\tilde\beta}  } \right)}}{{\Gamma \left( {\frac{1}{2} + {\tilde\beta}   - {\tilde\kappa} } \right)}} + {\mathcal{C}_2} \times {{\left( {2{\tilde\varepsilon} } \right)}^{ - {\tilde\kappa} }}{x^{ - 1 - {\tilde\kappa} }}\frac{{\Gamma \left( {1 - 2{\tilde\beta}  } \right)}}{{\Gamma \left( {\frac{1}{2} - {\tilde\beta}   - {\tilde\kappa} } \right)}}} \right]{e^{{\tilde\varepsilon} x}}.
\ee   
%%%%%%%%%%%%%%%%%%%%%%%%%%%%%%%%%%%%%%%%%%%%%%%%%%%%%%%%%%%%%%%
\section{Stationary Bound-State charged massive scalar fields}
\label{sec:bound-state}
The bound states of the charged massive scalar fields are portrayed by an exponentially decaying radial solution at infinity. This shows that the coefficient of the growing exponent ${e^{ \tilde\varepsilon x}}$ in (\ref{MR07}) must vanish,
\be \label{BS01}
{{\mathcal{C}_1}\times {{\left( {2\tilde\varepsilon } \right)}^{ - \tilde\kappa }}{x^{ - 1 - \tilde\kappa }}\frac{{\Gamma \left( {1 + 2\tilde\beta } \right)}}{{\Gamma \left( {\frac{1}{2} + \tilde\beta  - \tilde\kappa } \right)}} + {\mathcal{C}_2}\times {{\left( {2\tilde\varepsilon } \right)}^{ - \tilde\kappa }}{x^{ - 1 - \tilde\kappa }}\frac{{\Gamma \left( {1 - 2\tilde\beta } \right)}}{{\Gamma \left( {\frac{1}{2} - \tilde\beta  - \tilde\kappa } \right)}}}=0.
\ee
Substituting Eqs. (\ref{MR05}) and (\ref{MR06}) into (\ref{BS01}), one finds
\be \label{BS02}
\frac{1}{{\Gamma \left( {\frac{1}{2} + \tilde\beta  - \tilde\kappa } \right)}}= {\left( {2\tilde\varepsilon \tau } \right)^{2\tilde\beta }}{\left( {\frac{{\Gamma \left( { - 2\tilde\beta } \right)}}{{\Gamma \left( {2\tilde\beta } \right)}}} \right)^2}\frac{{\Gamma \left( {\frac{1}{2} + \tilde\beta  - i\tilde k} \right)\Gamma \left( {\frac{1}{2} + \tilde\beta  + i\tilde k} \right)}}{{\Gamma \left( {\frac{1}{2} - \tilde\beta  - i\tilde k } \right)\Gamma \left( {\frac{1}{2} - \tilde\beta  + i \tilde k } \right)\Gamma \left( {\frac{1}{2} - \tilde\beta  - \tilde\kappa } \right)}},
\ee
where $\tilde k = {2b{\omega _c} + k} $. The right-hand side of Eq. (\ref{BS02}) is of order $\mathcal{O}\left( {{{\left( {\varepsilon \tau } \right)}^{2\beta }}} \right) \ll 1$; therefore, Eq. (\ref{BS02}) can be composed within the form
\be \label{BS03}
\frac{1}{2} + \tilde\beta  - \tilde\kappa  = \mathcal{O}\left( {{{\left( {\tilde\varepsilon \tau } \right)}^{2\tilde\beta }}} \right) - {n},
\ee
with ${n} = 0,1,2,3, \ldots $. Equation (\ref{BS03}) can be solved within the regime
\be \label{BS04}
\tilde\varepsilon  \ll 1.
\ee
Taking note of Eqs. (\ref{SV06})--(\ref{SV09}) and (\ref{FR04})--(\ref{FR06}) within the regime (\ref{BS04}), one finds
\bea \label{BS05}
\tilde\beta  &=& {\tilde\beta _0} + \mathcal{O}\left( {{\tilde\varepsilon ^2}} \right),\\ \label{BS06}
\tilde\kappa  &=& \frac{\tilde\alpha }{\tilde\varepsilon } + \mathcal{O}\left( \tilde\varepsilon  \right),
\eea
where
\bea \label{BS07}
{\tilde\beta _0}
&\equiv& \sqrt {{{\left( {l + \frac{1}{2}} \right)}^2} - 2ma{\omega _c} - \varpi + \chi},\\ \label{BS08}
\tilde\alpha &\equiv& {r_ + }{\omega _c}\left( {{\omega _c}\left( {b + {r_ + }} \right) - qQ} \right), 
\eea
with $\varpi={{\left( {2{\omega _c}\left( {b + {r_ + }} \right) - qQ} \right)}^2}$ and $\chi={\omega _c}^2\left( {{r_ + }^2 + 2b{r_ + } + {a^2}} \right)$. Substituting Eqs. (\ref{BS05}) and (\ref{BS06}) into (\ref{BS03}), one finds
\be \label{BS09}
\tilde\varepsilon  = \frac{\tilde\alpha }{{\frac{1}{2} + {\tilde\beta _0} + n}}.
\ee
Taking note of Eqs. (\ref{BS05}), (\ref{BS06}), and (\ref{BS09}), one recognizes that $\tilde\alpha>0$ is a required condition for the existence of the stationary bound-state charged massive scalar fields. Finally, from Eq. (\ref{SV04}) the equations that are analogous to the stationary bound state of the massive charged scalar particle fields in Kerr-Sen spacetime are given by
\be \label{BS10}
\mu {r_ + } = \sqrt{{{\tilde\varepsilon} ^2} + {{\left( {{\omega _c}{r_ + }} \right)}^2}}. 
\ee
It turns out that Eq. (\ref{BS10}) for Kerr-Sen spacetime resembles the equation for Kerr-Newman spacetime as found by Hod in \cite{Hod:2014baa}. However, there are several distinguishable parameters between these two cases, namely, the explicit expressions for $\tilde\varepsilon$, $\tilde\beta_0$, and $\tilde\alpha$ that depend on $r_+$\footnote{For Kerr-Newman spacetime, ${r_ + } = M + \sqrt {{M^2} - {Q^2} - {a^2}}$, and for Kerr-Sen spacetime, ${r_ + } = \left( {M - b} \right) + \sqrt {\left( {{M^2} - {b^2}} \right) - {a^2}} $.}. 
%%%%%%%%%%%%%%%%%%%%%%%%%%%%%%%%%%%%%%%%%%%%%%%%%%%%%%%%%%%%%%%
\section{Charged Scalar Clouds in GMGHS Spacetime}
\label{sec:GMGHS}
A static spherical symmetric charged black hole in the low energy limit of heterotic string field theory in four dimensions was first found by Gibbons and Maeda in \cite{Gibbons:1987ps} and independently obtained by Garfinkle, Horowitz, and Strominger in \cite{Garfinkle:1990qj} three years later. The metric that describes this GMGHS spacetime is \cite{Li:2015bfa}
\be \label{SVGMGHS01}
d{s^2} =  - \left( {1 - \frac{{2M}}{r}} \right)d{t^2} + {\left( {1 - \frac{{2M}}{r}} \right)^{ - 1}}d{r^2} + r\left( {r - \frac{{{Q^2}}}{M}} \right)\left( {d{\theta ^2} + {{\sin }^2}d{\phi ^2}} \right).
\ee 
The corresponding potential vector and the dilaton field are
\bea \label{SVGMGHS02}
{A_t} &=&  - \frac{Q}{r},\\ \label{SVGMGHS03}
{e^{2\tilde \Phi }} &=& 1 - \frac{{{Q^2}}}{{Mr}}.
\eea
The event horizon of the GMGHS black hole is located at ${r_ + } = 2M$ and ${r_ - } = {Q^2}/M$. We start analyzing a massive charged scalar particle field $\Phi$ outside of a GMGHS black hole  with the mass $\mu$ and charge $q$ obeying the following Klein-Gordon wave equation: 
\be \label{SVGMGHS04}
\frac{1}{{\sqrt { - g} }}{\partial _\alpha }\left( {{g^{\alpha \beta }}\sqrt { - g} {\partial _\beta }\Phi } \right) - 2iq{A_\alpha }{g^{\alpha \beta }}{\partial _\beta }\Phi  - {q^2}{g^{\alpha \beta }}{A_\alpha }{A_\beta }\Phi  - {\mu ^2}\Phi  = 0.
\ee 
To solve the equation above, as usual we use the ansatz of the scalar field \cite{Degollado:2013eqa},
\be \label{SVGMGHS05}
\Phi=\sum\limits_{l,m}\Phi_{lm}  = \sum\limits_{l,m}{e^{ - i\omega t}}R_l\left( r \right){Y_{lm}}\left( {\theta ,\phi } \right),
\ee
where $l$ is the spherical harmonic index and $m$ is the azimuthal harmonic index with $- l \le m \le l$; one finds the radial equation \cite{Li:2013jna}
\be \label{SVGMGHS06}
{\Delta _{G}}\frac{d}{{dr}}\left( {{\Delta _{G}}\frac{dR_l\left( r \right)}{{dr}}} \right) + UR\left( r \right) = 0,
\ee
where
\be \label{SVGMGHS07}
{\Delta _G} = \left( {r - {r_ + }} \right)\left( {r - {r_ - }} \right),
\ee
and the potential $U$ is given by
\be \label{SVGMGHS08}
U = {\left( {r - \frac{{{Q^2}}}{M}} \right)^2}{\left( {\omega r - qQ} \right)^2} - {\Delta _G}\left[ {{\mu ^2}r\left( {r - \frac{{{Q^2}}}{M}} \right) + l\left( {l + 1} \right)} \right].
\ee 
\indent The charged massive scalar field with frequency $\omega=q\Phi_H$ represents the stationary bound state of the charged massive scalar field in GMGHS spacetime, where $\Phi_H=Q/2M$ is the electrostatic potential at the horizon \cite{Li:2015bfa}. Following \cite{Hod:2014baa,Li:2015bfa}, we outline the new dimensionless variables
\bea \label{SVGMGHS09}
x &\equiv& \frac{{r - {r_ + }}}{{{r_ + }}},\\\label{SVGMGHS10}
\tau  &\equiv& \frac{{{r_ + } - {r_ - }}}{{{r_ + }}},
\eea
in terms of which the radial equation (\ref{SVGMGHS06}) becomes
\be \label{SVGMGHS11}
x\left( {x + \tau } \right)\frac{{{d^{2}R\left( x \right)}}}{{d{x^2}}} + \left( {2x + \tau } \right)\frac{dR\left( x \right)}{{dx}} + \mathcal{U}R\left( x \right) = 0,
\ee
where
\be \label{SVGMGHS12}
\mathcal{U} = {q^2}{Q^2}x\left( {x + \tau } \right) - {\mu ^2}\left( {{r_ + }^2{{\left( {x + 1} \right)}^2} - 2{Q^2}\left( {x + 1} \right)} \right) - l\left( {l + 1} \right).
\ee
By setting $\mu=0$, i.e. a stationary massless charged scalar field, Eq. (\ref{SVGMGHS11}) becomes \cite{Li:2015bfa}
\be \label{SVGMGHS13}
x\left( {x + \tau } \right)\frac{{{d^{2}R\left( x \right)}}}{{d{x^2}}} + \left( {2x + \tau } \right)\frac{dR\left( x \right)}{{dx}} + \left[{q^2}{Q^2}x\left( {x + \tau } \right) - l\left( {l + 1} \right)\right]R\left( x \right) = 0.
\ee
Following Ref. \cite{Li:2015bfa}, Eq. (\ref{SVGMGHS13}) can be solved within the asymptotic highly charged regime $qQ\gg1$ \footnote{In this asymptotic highly charged regime, the charge $q$ of scalar field and the charge $Q$ of black hole are strongly interacting electrically.}, and within the near-horizon region $x\ll\tau$ \cite{Hod:2010hw,Hod:2012zzb,Konoplya:2013rxa,Hod:2014tqa}, one finds that the radial equation (\ref{SVGMGHS13}) can be approximated by
\be \label{SVGMGHS14}
x\frac{{{d^{2}R\left( x \right)}}}{{d{x^2}}} + \frac{dR\left( x \right)}{{dx}} + {q^2}{Q^2}x R\left( x \right) = 0.
\ee
The solution of the radial equation (\ref{SVGMGHS14}) is then given by the Bessel function of the first kind, 
\be \label{SVGMGHS15}
R\left( x \right) = {J_0}\left( {qQx} \right).
\ee
\indent The radial solution (\ref{SVGMGHS15}) describes the stationary scalar field around a static charged black hole in the low energy limit of heterotic string field theory, namely, the GMGHS black hole \cite{Li:2015bfa}.  This does not resembles the situation for a static charged black hole in Einstein-Maxwell theory, known as the Reissner-Nordstr\"{o}m black hole. In the Kerr-Newman black hole case ($a\ne0$) in Ref. \cite{Hod:2014baa}, setting the rotational parameter $a=0$ (a nonrotating charged  Reissner-Nordstr\"{o}m black hole), one finds that Eq. (29) in Ref. \cite{Hod:2014baa} has no solution and concludes that the Reissner-Nordstr\"{o}m black hole cannot support the existence of stationary scalar clouds \cite{Hod:2013eea}. However, setting the rotational parameter $a=0$ for the Kerr-Sen black hole, one finds $\tilde\alpha\ne0$; therefore, there is a solution for Eq. (\ref{BS03}). This is consistent with Li's works in \cite{Li:2015bfa} that concludes there is a stationary scalar field around charged black holes in the low energy limit of heterotic string field theory.
%%%%%%%%%%%%%%%%%%%%%%%%%%%%%%%%%%%%%%%%%%%%%%%%%%%%%%%%%%%%%%%
\section{Summary and Discussion}
\label{sec:summary}
In summary, we have studied the stationary massive charged scalar clouds in the Kerr-Sen black hole spacetime. In \cite{Hod:2014baa}, Hod showed that the Kerr-Newman black hole can support stationary massive charged scalar clouds by  analytically solving the Klein-Gordon equation for a stationary charged massive scalar fields. The authors of Ref. \cite{Huang:2016qnk} also numerically investigate stationary massive charged scalar clouds in the Kerr-Newman black hole spacetime and find that for fixed black hole parameters, the mass $\mu$ and charge $q$ of the scalar clouds are limited in a finite region in the parameter space of the scalar fields. The Kerr-Newman scalar clouds were recently promoted to a fully nonlinear solution (Kerr-Newman black holes with scalar hair) in Ref. \cite{Delgado:2016jxq}. Of course, the existence of clouds in this Kerr-Sen case shows that a new family of fully nonlinear solutions will also exist in the Kerr-Sen case. This motivated us to perform such an analysis for the Kerr-Sen black hole since Kerr-Newman and Kerr-Sen black holes have several similarities in their physical properties. In this paper, we have studied that the Kerr-Sen black hole can also support stationary massive charged scalar clouds.\\
\indent The stationary charged scalar clouds can be formed within the background of a static electrically charged black hole solution in the low energy limit of heterotic string field theory, namely, the GMGHS black holes at linear level: The gravitational attraction is equal because of the electromagnetic repulsion. The existence of this stationary scalar field was demonstrated numerically and analytically in the important work by Li, Zhao, Wu, and Zhang \cite{Li:2015bfa}. This does not resemble the situation in Einstein-Maxwell theory, where the Reissner-Nordstr\"om black holes cannot support the existence of stationary scalar clouds because the gravitational attraction and electromagnetic repulsion cannot reach equilibrium \cite{Hod:2013eea,Degollado:2013eqa,Li:2015bfa}. Lastly, we note that it is worth performing a numerical analysis of the linear stationary charged scalar clouds for the rotating and charged black hole in the low energy limit of heterotic string field theory. Intuitively, we predict that the same result can be found for the stationary scalar clouds surrounding a Kerr-Sen black hole. There might be related numerical works, but we hypothesize that the situation is not exactly the same; it can be found in Refs. \cite{Hwang:2011mn,Hansen:2013vha}. Perhaps some of these results would approach some of stationary solutions in this paper, though we are not completely sure. For this case, there seems to be scalar or charge clouds because of AdS background. In Refs. \cite{Hansen:2014rua,Hansen:2015dxa,Nakonieczna:2015umf,Nakonieczna:2016iof} we found various string-inspired models that allow various scalar hairs (normal scalar as well as Brans-Dicke scalar). In this case, a hair appears due to a special coupling with the electric charge.
%%%%%%%%%%%%%%%%%%%%%%%%%%%%%%%%%%%%%%%%%%%%%%%%%%%%%%%%%%%%%%%
\section*{Acknowledgments}
I am very thankful to my supervisor Haryanto M.~Siahaan for all his support and the knowledge he shared with me. I also thank Professor Carlos A. R. Herdeiro and the anonymous referee for reading this manuscript and for their useful comments and suggestions.
%%%%%%%%%%%%%%%%%%%%%%%%%%%%%%%%%%%%%%%%%%%%%%%%%%%%%%%%%%%%%%%
\bibliography{kssc}

\begin{thebibliography}{35}%
\makeatletter
\providecommand \@ifxundefined [1]{%
 \@ifx{#1\undefined}
}%
\providecommand \@ifnum [1]{%
 \ifnum #1\expandafter \@firstoftwo
 \else \expandafter \@secondoftwo
 \fi
}%
\providecommand \@ifx [1]{%
 \ifx #1\expandafter \@firstoftwo
 \else \expandafter \@secondoftwo
 \fi
}%
\providecommand \natexlab [1]{#1}%
\providecommand \enquote  [1]{``#1''}%
\providecommand \bibnamefont  [1]{#1}%
\providecommand \bibfnamefont [1]{#1}%
\providecommand \citenamefont [1]{#1}%
\providecommand \href@noop [0]{\@secondoftwo}%
\providecommand \href [0]{\begingroup \@sanitize@url \@href}%
\providecommand \@href[1]{\@@startlink{#1}\@@href}%
\providecommand \@@href[1]{\endgroup#1\@@endlink}%
\providecommand \@sanitize@url [0]{\catcode `\\12\catcode `\$12\catcode
  `\&12\catcode `\#12\catcode `\^12\catcode `\_12\catcode `\%12\relax}%
\providecommand \@@startlink[1]{}%
\providecommand \@@endlink[0]{}%
\providecommand \url  [0]{\begingroup\@sanitize@url \@url }%
\providecommand \@url [1]{\endgroup\@href {#1}{\urlprefix }}%
\providecommand \urlprefix  [0]{URL }%
\providecommand \Eprint [0]{\href }%
\providecommand \doibase [0]{http://dx.doi.org/}%
\providecommand \selectlanguage [0]{\@gobble}%
\providecommand \bibinfo  [0]{\@secondoftwo}%
\providecommand \bibfield  [0]{\@secondoftwo}%
\providecommand \translation [1]{[#1]}%
\providecommand \BibitemOpen [0]{}%
\providecommand \bibitemStop [0]{}%
\providecommand \bibitemNoStop [0]{.\EOS\space}%
\providecommand \EOS [0]{\spacefactor3000\relax}%
\providecommand \BibitemShut  [1]{\csname bibitem#1\endcsname}%
\let\auto@bib@innerbib\@empty
%</preamble>
\bibitem [{\citenamefont {Chrusciel}\ \emph {et~al.}(2012)\citenamefont
  {Chrusciel}, \citenamefont {Lopes~Costa},\ and\ \citenamefont
  {Heusler}}]{Chrusciel:2012jk}%
  \BibitemOpen
  \bibfield  {author} {\bibinfo {author} {\bibfnamefont {P.~T.}\ \bibnamefont
  {Chrusciel}}, \bibinfo {author} {\bibfnamefont {J.}~\bibnamefont
  {Lopes~Costa}}, \ and\ \bibinfo {author} {\bibfnamefont {M.}~\bibnamefont
  {Heusler}},\ }\href {\doibase 10.12942/lrr-2012-7} {\bibfield  {journal}
  {\bibinfo  {journal} {Living Rev. Rel.}\ }\textbf {\bibinfo {volume} {15}},\
  \bibinfo {pages} {7} (\bibinfo {year} {2012})}\BibitemShut {NoStop}%
\bibitem [{\citenamefont {Herdeiro}\ and\ \citenamefont
  {Radu}(2015{\natexlab{a}})}]{Herdeiro:2015waa}%
  \BibitemOpen
  \bibfield  {author} {\bibinfo {author} {\bibfnamefont {C.~A.~R.}\
  \bibnamefont {Herdeiro}}\ and\ \bibinfo {author} {\bibfnamefont
  {E.}~\bibnamefont {Radu}},\ }\href {\doibase 10.1142/S0218271815420146}
  {\bibfield  {journal} {\bibinfo  {journal} {Int. J. Mod. Phys.}\ }\textbf
  {\bibinfo {volume} {D 24}},\ \bibinfo {pages} {1542014} (\bibinfo {year}
  {2015}{\natexlab{a}})}\BibitemShut {NoStop}%
\bibitem [{\citenamefont {Sen}(1992)}]{Sen:1992ua}%
  \BibitemOpen
  \bibfield  {author} {\bibinfo {author} {\bibfnamefont {A.}~\bibnamefont
  {Sen}},\ }\href {\doibase 10.1103/PhysRevLett.69.1006} {\bibfield  {journal}
  {\bibinfo  {journal} {Phys. Rev. Lett.}\ }\textbf {\bibinfo {volume} {69}},\
  \bibinfo {pages} {1006} (\bibinfo {year} {1992})}\BibitemShut {NoStop}%
\bibitem [{\citenamefont {Hod}(2014{\natexlab{a}})}]{Hod:2014baa}%
  \BibitemOpen
  \bibfield  {author} {\bibinfo {author} {\bibfnamefont {S.}~\bibnamefont
  {Hod}},\ }\href {\doibase 10.1103/PhysRevD.90.024051} {\bibfield  {journal}
  {\bibinfo  {journal} {Phys. Rev.}\ }\textbf {\bibinfo {volume} {D 90}},\
  \bibinfo {pages} {024051} (\bibinfo {year} {2014}{\natexlab{a}})}\BibitemShut
  {NoStop}%
\bibitem [{\citenamefont {Herdeiro}\ and\ \citenamefont
  {Radu}(2014{\natexlab{a}})}]{Herdeiro:2014goa}%
  \BibitemOpen
  \bibfield  {author} {\bibinfo {author} {\bibfnamefont {C.~A.~R.}\
  \bibnamefont {Herdeiro}}\ and\ \bibinfo {author} {\bibfnamefont
  {E.}~\bibnamefont {Radu}},\ }\href {\doibase 10.1103/PhysRevLett.112.221101}
  {\bibfield  {journal} {\bibinfo  {journal} {Phys. Rev. Lett.}\ }\textbf
  {\bibinfo {volume} {112}},\ \bibinfo {pages} {221101} (\bibinfo {year}
  {2014}{\natexlab{a}})}\BibitemShut {NoStop}%
\bibitem [{\citenamefont {Herdeiro}\ and\ \citenamefont
  {Radu}(2015{\natexlab{b}})}]{Herdeiro:2015gia}%
  \BibitemOpen
  \bibfield  {author} {\bibinfo {author} {\bibfnamefont {C.}~\bibnamefont
  {Herdeiro}}\ and\ \bibinfo {author} {\bibfnamefont {E.}~\bibnamefont
  {Radu}},\ }\href {\doibase 10.1088/0264-9381/32/14/144001} {\bibfield
  {journal} {\bibinfo  {journal} {Class. Quant. Grav.}\ }\textbf {\bibinfo
  {volume} {32}},\ \bibinfo {pages} {144001} (\bibinfo {year}
  {2015}{\natexlab{b}})}\BibitemShut {NoStop}%
\bibitem [{\citenamefont {Herdeiro}\ and\ \citenamefont
  {Radu}(2014{\natexlab{b}})}]{Herdeiro:2014ima}%
  \BibitemOpen
  \bibfield  {author} {\bibinfo {author} {\bibfnamefont {C.~A.~R.}\
  \bibnamefont {Herdeiro}}\ and\ \bibinfo {author} {\bibfnamefont
  {E.}~\bibnamefont {Radu}},\ }\href {\doibase 10.1142/S0218271814420140}
  {\bibfield  {journal} {\bibinfo  {journal} {Int. J. Mod. Phys.}\ }\textbf
  {\bibinfo {volume} {D 23}},\ \bibinfo {pages} {1442014} (\bibinfo {year}
  {2014}{\natexlab{b}})}\BibitemShut {NoStop}%
\bibitem [{\citenamefont {Benone}\ \emph {et~al.}(2014)\citenamefont {Benone},
  \citenamefont {Crispino}, \citenamefont {Herdeiro},\ and\ \citenamefont
  {Radu}}]{Benone:2014ssa}%
  \BibitemOpen
  \bibfield  {author} {\bibinfo {author} {\bibfnamefont {C.~L.}\ \bibnamefont
  {Benone}}, \bibinfo {author} {\bibfnamefont {L.~C.~B.}\ \bibnamefont
  {Crispino}}, \bibinfo {author} {\bibfnamefont {C.}~\bibnamefont {Herdeiro}},
  \ and\ \bibinfo {author} {\bibfnamefont {E.}~\bibnamefont {Radu}},\ }\href
  {\doibase 10.1103/PhysRevD.90.104024} {\bibfield  {journal} {\bibinfo
  {journal} {Phys. Rev.}\ }\textbf {\bibinfo {volume} {D 90}},\ \bibinfo
  {pages} {104024} (\bibinfo {year} {2014})}\BibitemShut {NoStop}%
\bibitem [{\citenamefont {Siahaan}(2015)}]{Siahaan:2015xna}%
  \BibitemOpen
  \bibfield  {author} {\bibinfo {author} {\bibfnamefont {H.~M.}\ \bibnamefont
  {Siahaan}},\ }\href {\doibase 10.1142/S0218271815501023} {\bibfield
  {journal} {\bibinfo  {journal} {Int. J. Mod. Phys.}\ }\textbf {\bibinfo
  {volume} {D 24}},\ \bibinfo {pages} {1550102} (\bibinfo {year}
  {2015})}\BibitemShut {NoStop}%
\bibitem [{\citenamefont {Ghezelbash}\ and\ \citenamefont
  {Siahaan}(2013)}]{Ghezelbash:2012qn}%
  \BibitemOpen
  \bibfield  {author} {\bibinfo {author} {\bibfnamefont {A.~M.}\ \bibnamefont
  {Ghezelbash}}\ and\ \bibinfo {author} {\bibfnamefont {H.~M.}\ \bibnamefont
  {Siahaan}},\ }\href {\doibase 10.1088/0264-9381/30/13/135005} {\bibfield
  {journal} {\bibinfo  {journal} {Class. Quant. Grav.}\ }\textbf {\bibinfo
  {volume} {30}},\ \bibinfo {pages} {135005} (\bibinfo {year}
  {2013})}\BibitemShut {NoStop}%
\bibitem [{\citenamefont {Siahaan}(2016)}]{Siahaan:2015ljs}%
  \BibitemOpen
  \bibfield  {author} {\bibinfo {author} {\bibfnamefont {H.~M.}\ \bibnamefont
  {Siahaan}},\ }\href {\doibase 10.1103/PhysRevD.93.064028} {\bibfield
  {journal} {\bibinfo  {journal} {Phys. Rev.}\ }\textbf {\bibinfo {volume} {D
  93}},\ \bibinfo {pages} {064028} (\bibinfo {year} {2016})}\BibitemShut
  {NoStop}%
\bibitem [{\citenamefont {Wu}\ and\ \citenamefont {Cai}(2003)}]{Wu:2003qb}%
  \BibitemOpen
  \bibfield  {author} {\bibinfo {author} {\bibfnamefont {S.~Q.}\ \bibnamefont
  {Wu}}\ and\ \bibinfo {author} {\bibfnamefont {X.}~\bibnamefont {Cai}},\
  }\href {\doibase 10.1063/1.1539899} {\bibfield  {journal} {\bibinfo
  {journal} {J. Math. Phys.}\ }\textbf {\bibinfo {volume} {44}},\ \bibinfo
  {pages} {1084} (\bibinfo {year} {2003})}\BibitemShut {NoStop}%
\bibitem [{\citenamefont {Gibbons}\ and\ \citenamefont
  {Maeda}(1988)}]{Gibbons:1987ps}%
  \BibitemOpen
  \bibfield  {author} {\bibinfo {author} {\bibfnamefont {G.~W.}\ \bibnamefont
  {Gibbons}}\ and\ \bibinfo {author} {\bibfnamefont {K.-i.}\ \bibnamefont
  {Maeda}},\ }\href {\doibase 10.1016/0550-3213(88)90006-5} {\bibfield
  {journal} {\bibinfo  {journal} {Nucl. Phys.}\ }\textbf {\bibinfo {volume}
  {B298}},\ \bibinfo {pages} {741} (\bibinfo {year} {1988})}\BibitemShut
  {NoStop}%
%%CITATION = NUPHA,B298,741;%%
\bibitem [{\citenamefont {Garfinkle}\ \emph {et~al.}(1991)\citenamefont
  {Garfinkle}, \citenamefont {Horowitz},\ and\ \citenamefont
  {Strominger}}]{Garfinkle:1990qj}%
  \BibitemOpen
  \bibfield  {author} {\bibinfo {author} {\bibfnamefont {D.}~\bibnamefont
  {Garfinkle}}, \bibinfo {author} {\bibfnamefont {G.~T.}\ \bibnamefont
  {Horowitz}}, \ and\ \bibinfo {author} {\bibfnamefont {A.}~\bibnamefont
  {Strominger}},\ }\href {\doibase 10.1103/PhysRevD.43.3140,
  10.1103/PhysRevD.45.3888} {\bibfield  {journal} {\bibinfo  {journal} {Phys.
  Rev.}\ }\textbf {\bibinfo {volume} {D 43}},\ \bibinfo {pages} {3140}
  (\bibinfo {year} {1991})},\ \bibinfo {note} {[Erratum: Phys. Rev. D 45, 3888
  (1992)]}\BibitemShut {NoStop}%
%%CITATION = PHRVA,D43,3140;%%
\bibitem [{\citenamefont {Konoplya}\ and\ \citenamefont
  {Zhidenko}(2013)}]{Konoplya:2013rxa}%
  \BibitemOpen
  \bibfield  {author} {\bibinfo {author} {\bibfnamefont {R.~A.}\ \bibnamefont
  {Konoplya}}\ and\ \bibinfo {author} {\bibfnamefont {A.}~\bibnamefont
  {Zhidenko}},\ }\href {\doibase 10.1103/PhysRevD.88.024054} {\bibfield
  {journal} {\bibinfo  {journal} {Phys. Rev.}\ }\textbf {\bibinfo {volume} {D
  88}},\ \bibinfo {pages} {024054} (\bibinfo {year} {2013})}\BibitemShut
  {NoStop}%
\bibitem [{\citenamefont {Brill}\ \emph {et~al.}(1972)\citenamefont {Brill},
  \citenamefont {Chrzanowski}, \citenamefont {Martin~Pereira}, \citenamefont
  {Fackerell},\ and\ \citenamefont {Ipser}}]{Brill:1972xj}%
  \BibitemOpen
  \bibfield  {author} {\bibinfo {author} {\bibfnamefont {D.~R.}\ \bibnamefont
  {Brill}}, \bibinfo {author} {\bibfnamefont {P.~L.}\ \bibnamefont
  {Chrzanowski}}, \bibinfo {author} {\bibfnamefont {C.}~\bibnamefont
  {Martin~Pereira}}, \bibinfo {author} {\bibfnamefont {E.~D.}\ \bibnamefont
  {Fackerell}}, \ and\ \bibinfo {author} {\bibfnamefont {J.~R.}\ \bibnamefont
  {Ipser}},\ }\href {\doibase 10.1103/PhysRevD.5.1913} {\bibfield  {journal}
  {\bibinfo  {journal} {Phys. Rev.}\ }\textbf {\bibinfo {volume} {D 5}},\
  \bibinfo {pages} {1913} (\bibinfo {year} {1972})}\BibitemShut {NoStop}%
%%CITATION = PHRVA,D5,1913;%%
\bibitem [{\citenamefont {Dolan}(2007)}]{Dolan:2007mj}%
  \BibitemOpen
  \bibfield  {author} {\bibinfo {author} {\bibfnamefont {S.~R.}\ \bibnamefont
  {Dolan}},\ }\href {\doibase 10.1103/PhysRevD.76.084001} {\bibfield  {journal}
  {\bibinfo  {journal} {Phys. Rev.}\ }\textbf {\bibinfo {volume} {D 76}},\
  \bibinfo {pages} {084001} (\bibinfo {year} {2007})}\BibitemShut {NoStop}%
\bibitem [{\citenamefont {Abramowitz}\ and\ \citenamefont
  {Stegun}(1970)}]{Abra}%
  \BibitemOpen
  \bibfield  {author} {\bibinfo {author} {\bibfnamefont {M.}~\bibnamefont
  {Abramowitz}}\ and\ \bibinfo {author} {\bibfnamefont {I.~A.}\ \bibnamefont
  {Stegun}},\ }\href@noop {} {\emph {\bibinfo {title} {Handbook of Mathematical
  Functions}}},\ \bibinfo {edition} {9th}\ ed.\ (\bibinfo  {publisher} {Dover
  Publication, New York},\ \bibinfo {year} {1970})\ pp.\ \bibinfo {pages}
  {255--566}\BibitemShut {NoStop}%
\bibitem [{\citenamefont {Hartman}\ \emph {et~al.}(2010)\citenamefont
  {Hartman}, \citenamefont {Song},\ and\ \citenamefont
  {Strominger}}]{Hartman:2009nz}%
  \BibitemOpen
  \bibfield  {author} {\bibinfo {author} {\bibfnamefont {T.}~\bibnamefont
  {Hartman}}, \bibinfo {author} {\bibfnamefont {W.}~\bibnamefont {Song}}, \
  and\ \bibinfo {author} {\bibfnamefont {A.}~\bibnamefont {Strominger}},\
  }\href {\doibase 10.1007/JHEP03(2010)118} {\bibfield  {journal} {\bibinfo
  {journal} {JHEP}\ }\textbf {\bibinfo {volume} {03}},\ \bibinfo {pages} {118}
  (\bibinfo {year} {2010})}\BibitemShut {NoStop}%
\bibitem [{\citenamefont {Hod}(2012{\natexlab{a}})}]{Hod:2012px}%
  \BibitemOpen
  \bibfield  {author} {\bibinfo {author} {\bibfnamefont {S.}~\bibnamefont
  {Hod}},\ }\href {\doibase 10.1103/PhysRevD.86.129902,
  10.1103/PhysRevD.86.104026} {\bibfield  {journal} {\bibinfo  {journal} {Phys.
  Rev.}\ }\textbf {\bibinfo {volume} {D 86}},\ \bibinfo {pages} {104026}
  (\bibinfo {year} {2012}{\natexlab{a}})},\ \bibinfo {note} {[Erratum: Phys.
  Rev. D 86, 129902 (2012)]}\BibitemShut {NoStop}%
\bibitem [{\citenamefont {Li}\ \emph {et~al.}(2015)\citenamefont {Li},
  \citenamefont {Zhao}, \citenamefont {Wu},\ and\ \citenamefont
  {Zhang}}]{Li:2015bfa}%
  \BibitemOpen
  \bibfield  {author} {\bibinfo {author} {\bibfnamefont {R.}~\bibnamefont
  {Li}}, \bibinfo {author} {\bibfnamefont {J.}~\bibnamefont {Zhao}}, \bibinfo
  {author} {\bibfnamefont {X.}~\bibnamefont {Wu}}, \ and\ \bibinfo {author}
  {\bibfnamefont {Y.}~\bibnamefont {Zhang}},\ }\href {\doibase
  10.1140/epjc/s10052-015-3370-4} {\bibfield  {journal} {\bibinfo  {journal}
  {Eur. Phys. J.}\ }\textbf {\bibinfo {volume} {C 75}},\ \bibinfo {pages} {142}
  (\bibinfo {year} {2015})}\BibitemShut {NoStop}%
\bibitem [{\citenamefont {Degollado}\ and\ \citenamefont
  {Herdeiro}(2013)}]{Degollado:2013eqa}%
  \BibitemOpen
  \bibfield  {author} {\bibinfo {author} {\bibfnamefont {J.~C.}\ \bibnamefont
  {Degollado}}\ and\ \bibinfo {author} {\bibfnamefont {C.~A.~R.}\ \bibnamefont
  {Herdeiro}},\ }\href {\doibase 10.1007/s10714-013-1598-6} {\bibfield
  {journal} {\bibinfo  {journal} {Gen. Rel. Grav.}\ }\textbf {\bibinfo {volume}
  {45}},\ \bibinfo {pages} {2483} (\bibinfo {year} {2013})}\BibitemShut
  {NoStop}%
\bibitem [{\citenamefont {Li}(2013)}]{Li:2013jna}%
  \BibitemOpen
  \bibfield  {author} {\bibinfo {author} {\bibfnamefont {R.}~\bibnamefont
  {Li}},\ }\href {\doibase 10.1103/PhysRevD.88.127901} {\bibfield  {journal}
  {\bibinfo  {journal} {Phys. Rev.}\ }\textbf {\bibinfo {volume} {D 88}},\
  \bibinfo {pages} {127901} (\bibinfo {year} {2013})}\BibitemShut {NoStop}%
\bibitem [{\citenamefont {Hod}(2010)}]{Hod:2010hw}%
  \BibitemOpen
  \bibfield  {author} {\bibinfo {author} {\bibfnamefont {S.}~\bibnamefont
  {Hod}},\ }\href {\doibase 10.1016/j.physleta.2010.05.052} {\bibfield
  {journal} {\bibinfo  {journal} {Phys. Lett.}\ }\textbf {\bibinfo {volume} {A
  374}},\ \bibinfo {pages} {2901} (\bibinfo {year} {2010})}\BibitemShut
  {NoStop}%
\bibitem [{\citenamefont {Hod}(2012{\natexlab{b}})}]{Hod:2012zzb}%
  \BibitemOpen
  \bibfield  {author} {\bibinfo {author} {\bibfnamefont {S.}~\bibnamefont
  {Hod}},\ }\href {\doibase 10.1016/j.physletb.2012.03.010} {\bibfield
  {journal} {\bibinfo  {journal} {Phys. Lett.}\ }\textbf {\bibinfo {volume} {B
  710}},\ \bibinfo {pages} {349} (\bibinfo {year}
  {2012}{\natexlab{b}})}\BibitemShut {NoStop}%
\bibitem [{\citenamefont {Hod}(2014{\natexlab{b}})}]{Hod:2014tqa}%
  \BibitemOpen
  \bibfield  {author} {\bibinfo {author} {\bibfnamefont {S.}~\bibnamefont
  {Hod}},\ }\href {\doibase 10.1140/epjc/s10052-014-3137-3} {\bibfield
  {journal} {\bibinfo  {journal} {Eur. Phys. J.}\ }\textbf {\bibinfo {volume}
  {C 74}},\ \bibinfo {pages} {3137} (\bibinfo {year}
  {2014}{\natexlab{b}})}\BibitemShut {NoStop}%
\bibitem [{\citenamefont {Hod}(2012{\natexlab{c}})}]{Hod:2013eea}%
  \BibitemOpen
  \bibfield  {author} {\bibinfo {author} {\bibfnamefont {S.}~\bibnamefont
  {Hod}},\ }\href {\doibase 10.1016/j.physletb.2012.06.043} {\bibfield
  {journal} {\bibinfo  {journal} {Phys. Lett.}\ }\textbf {\bibinfo {volume} {B
  713}},\ \bibinfo {pages} {505} (\bibinfo {year}
  {2012}{\natexlab{c}})}\BibitemShut {NoStop}%
\bibitem [{\citenamefont {Huang}\ and\ \citenamefont
  {Liu}(2016)}]{Huang:2016qnk}%
  \BibitemOpen
  \bibfield  {author} {\bibinfo {author} {\bibfnamefont {Y.}~\bibnamefont
  {Huang}}\ and\ \bibinfo {author} {\bibfnamefont {D.-J.}\ \bibnamefont
  {Liu}},\ }\href {\doibase 10.1103/PhysRevD.94.064030} {\bibfield  {journal}
  {\bibinfo  {journal} {Phys. Rev.}\ }\textbf {\bibinfo {volume} {D 94}},\
  \bibinfo {pages} {064030} (\bibinfo {year} {2016})}\BibitemShut {NoStop}%
\bibitem [{\citenamefont {Delgado}\ \emph {et~al.}(2016)\citenamefont
  {Delgado}, \citenamefont {Herdeiro}, \citenamefont {Radu},\ and\
  \citenamefont {Runarsson}}]{Delgado:2016jxq}%
  \BibitemOpen
  \bibfield  {author} {\bibinfo {author} {\bibfnamefont {J.~F.~M.}\
  \bibnamefont {Delgado}}, \bibinfo {author} {\bibfnamefont {C.~A.~R.}\
  \bibnamefont {Herdeiro}}, \bibinfo {author} {\bibfnamefont {E.}~\bibnamefont
  {Radu}}, \ and\ \bibinfo {author} {\bibfnamefont {H.}~\bibnamefont
  {Runarsson}},\ }\href {\doibase 10.1016/j.physletb.2016.08.032} {\bibfield
  {journal} {\bibinfo  {journal} {Phys. Lett.}\ }\textbf {\bibinfo {volume} {B
  761}},\ \bibinfo {pages} {234} (\bibinfo {year} {2016})}\BibitemShut
  {NoStop}%
\bibitem [{\citenamefont {Hwang}\ \emph {et~al.}(2012)\citenamefont {Hwang},
  \citenamefont {Kim},\ and\ \citenamefont {Yeom}}]{Hwang:2011mn}%
  \BibitemOpen
  \bibfield  {author} {\bibinfo {author} {\bibfnamefont {D.-i.}\ \bibnamefont
  {Hwang}}, \bibinfo {author} {\bibfnamefont {H.-B.}\ \bibnamefont {Kim}}, \
  and\ \bibinfo {author} {\bibfnamefont {D.-h.}\ \bibnamefont {Yeom}},\ }\href
  {\doibase 10.1088/0264-9381/29/5/055003} {\bibfield  {journal} {\bibinfo
  {journal} {Class. Quant. Grav.}\ }\textbf {\bibinfo {volume} {29}},\ \bibinfo
  {pages} {055003} (\bibinfo {year} {2012})}\BibitemShut {NoStop}%
\bibitem [{\citenamefont {Hansen}\ \emph {et~al.}(2013)\citenamefont {Hansen},
  \citenamefont {Lee}, \citenamefont {Park},\ and\ \citenamefont
  {Yeom}}]{Hansen:2013vha}%
  \BibitemOpen
  \bibfield  {author} {\bibinfo {author} {\bibfnamefont {J.}~\bibnamefont
  {Hansen}}, \bibinfo {author} {\bibfnamefont {B.-H.}\ \bibnamefont {Lee}},
  \bibinfo {author} {\bibfnamefont {C.}~\bibnamefont {Park}}, \ and\ \bibinfo
  {author} {\bibfnamefont {D.-h.}\ \bibnamefont {Yeom}},\ }\href {\doibase
  10.1088/0264-9381/30/23/235022} {\bibfield  {journal} {\bibinfo  {journal}
  {Class. Quant. Grav.}\ }\textbf {\bibinfo {volume} {30}},\ \bibinfo {pages}
  {235022} (\bibinfo {year} {2013})}\BibitemShut {NoStop}%
\bibitem [{\citenamefont {Hansen}\ and\ \citenamefont
  {Yeom}(2014)}]{Hansen:2014rua}%
  \BibitemOpen
  \bibfield  {author} {\bibinfo {author} {\bibfnamefont {J.}~\bibnamefont
  {Hansen}}\ and\ \bibinfo {author} {\bibfnamefont {D.-h.}\ \bibnamefont
  {Yeom}},\ }\href {\doibase 10.1007/JHEP10(2014)040} {\bibfield  {journal}
  {\bibinfo  {journal} {JHEP}\ }\textbf {\bibinfo {volume} {10}},\ \bibinfo
  {pages} {040} (\bibinfo {year} {2014})}\BibitemShut {NoStop}%
\bibitem [{\citenamefont {Hansen}\ and\ \citenamefont
  {Yeom}(2015)}]{Hansen:2015dxa}%
  \BibitemOpen
  \bibfield  {author} {\bibinfo {author} {\bibfnamefont {J.}~\bibnamefont
  {Hansen}}\ and\ \bibinfo {author} {\bibfnamefont {D.-h.}\ \bibnamefont
  {Yeom}},\ }\href {\doibase 10.1088/1475-7516/2015/09/019} {\bibfield
  {journal} {\bibinfo  {journal} {JCAP}\ }\textbf {\bibinfo {volume} {1509}},\
  \bibinfo {pages} {019} (\bibinfo {year} {2015})}\BibitemShut {NoStop}%
\bibitem [{\citenamefont {Nakonieczna}\ and\ \citenamefont
  {Yeom}(2016{\natexlab{a}})}]{Nakonieczna:2015umf}%
  \BibitemOpen
  \bibfield  {author} {\bibinfo {author} {\bibfnamefont {A.}~\bibnamefont
  {Nakonieczna}}\ and\ \bibinfo {author} {\bibfnamefont {D.-h.}\ \bibnamefont
  {Yeom}},\ }\href {\doibase 10.1007/JHEP02(2016)049} {\bibfield  {journal}
  {\bibinfo  {journal} {JHEP}\ }\textbf {\bibinfo {volume} {02}},\ \bibinfo
  {pages} {049} (\bibinfo {year} {2016}{\natexlab{a}})}\BibitemShut {NoStop}%
\bibitem [{\citenamefont {Nakonieczna}\ and\ \citenamefont
  {Yeom}(2016{\natexlab{b}})}]{Nakonieczna:2016iof}%
  \BibitemOpen
  \bibfield  {author} {\bibinfo {author} {\bibfnamefont {A.}~\bibnamefont
  {Nakonieczna}}\ and\ \bibinfo {author} {\bibfnamefont {D.-h.}\ \bibnamefont
  {Yeom}},\ }\href {\doibase 10.1007/JHEP05(2016)155} {\bibfield  {journal}
  {\bibinfo  {journal} {JHEP}\ }\textbf {\bibinfo {volume} {05}},\ \bibinfo
  {pages} {155} (\bibinfo {year} {2016}{\natexlab{b}})}\BibitemShut {NoStop}%
\end{thebibliography}%
\bibliographystyle{apsrev4-1}
\end{document}